\documentclass[11pt,twocolumn]{aastex63}

\usepackage{mathtools}   
\usepackage{makecell}

\definecolor{myblue}{RGB}{15, 82, 136}

\shorttitle{Two Rings and a Marginally Resolved Disk Around LkCa\,15}
\shortauthors{Blakely et al.}

\graphicspath{{./}{Figures/}}

\begin{document}

\title{Two Rings and a Marginally Resolved, 5\,AU, Disk Around LkCa\,15 Identified Via Near Infrared Sparse Aperture Masking Interferometry}\footnote{Released on XX, XX, XXXX}

\author[0000-0001-9582-4261]{Dori Blakely}
\affiliation{Department of Physics and Astronomy, University of Victoria,
3800 Finnerty Road, Elliot Building,
Victoria, BC, V8P 5C2, Canada}
\affiliation{NRC Herzberg Astronomy and Astrophysics,
5071 West Saanich Road,
Victoria, BC, V9E 2E7, Canada}

\author[0000-0001-8822-6327]{Logan Francis}
\affiliation{Department of Physics and Astronomy, University of Victoria,
3800 Finnerty Road, Elliot Building,
Victoria, BC, V8P 5C2, Canada}
\affiliation{NRC Herzberg Astronomy and Astrophysics,
5071 West Saanich Road,
Victoria, BC, V9E 2E7, Canada}

\author[0000-0002-6773-459X]{Doug Johnstone}
\affiliation{NRC Herzberg Astronomy and Astrophysics,
5071 West Saanich Road,
Victoria, BC, V9E 2E7, Canada}
\affiliation{Department of Physics and Astronomy, University of Victoria,
3800 Finnerty Road, Elliot Building,
Victoria, BC, V8P 5C2, Canada}

\author{Anthony Soulain}
\affiliation{Sydney Institute for Astronomy, School of Physics,
University of Sydney, NSW 2006, Australia}
\affiliation{Univ.\ Grenoble Alpes, CNRS, IPAG, 38100 Grenoble, France}

\author{Peter Tuthill }
\affiliation{Sydney Institute for Astronomy, School of Physics,
University of Sydney, NSW 2006, Australia}

\author{Anthony Cheetham}
\affiliation{Sydney Institute for Astronomy, School of Physics,
University of Sydney, NSW 2006, Australia}

\author{Joel Sanchez-Bermudez}
\affiliation{Instituto de Astronomía, Universidad Nacional Autónoma de México, Apdo. Postal 70264, Ciudad de México, 04510, México}
\affiliation{Max-Planck-Institut für Astronomie, Königstuhl 17, D-69117 Heidelberg, Germany}

\author[0000-0003-1251-4124]{Anand Sivaramakrishnan}
\affiliation{Space Telescope Science Institute, 3700 San Martin Drive, Baltimore, MD 21218, USA}
\affiliation{Johns Hopkins University Department of Physics and Astronomy 3400 North Charles, Baltimore, MD 21218}

\author[0000-0001-9290-7846]{Ruobing Dong}
\affiliation{Department of Physics and Astronomy, University of Victoria,
3800 Finnerty Road, Elliot Building,
Victoria, BC, V8P 5C2, Canada}

\author[0000-0003-2458-9756]{Nienke van der Marel}
\affiliation{Department of Physics and Astronomy, University of Victoria,
3800 Finnerty Road, Elliot Building,
Victoria, BC, V8P 5C2, Canada}
\affiliation{Leiden Observatory, Leiden University, 2300 RA, Leiden, The Netherlands}

\author[0000-0001-7864-308X]{Rachel Cooper}
\affiliation{Space Telescope Science Institute, 3700 San Martin Drive, Baltimore, MD 21218, USA}

\author{Arthur Vigan}
\affiliation{Aix Marseille Univ, CNRS, CNES, LAM, Marseille, France}

\author{Faustine Cantalloube}
\affiliation{Aix Marseille Univ, CNRS, CNES, LAM, Marseille, France}




\begin{abstract}
Sparse aperture masking interferometry (SAM) is a high resolution observing technique that allows for imaging at and beyond a telescope's diffraction limit. The technique is ideal for searching for stellar companions at small separations from their host star; however, previous analysis of SAM observations of young stars surrounded by dusty disks have had difficulties disentangling planet and extended disk emission. We analyse VLT/SPHERE-IRDIS SAM observations of the transition disk LkCa\,15, model the extended disk emission, probe for planets at small separations, and improve contrast limits for planets. We fit geometrical models directly to the interferometric observables and recover previously observed extended disk emission. We use dynamic nested sampling to estimate uncertainties on our model parameters and to calculate evidences to perform model comparison. We compare our extended disk emission models against point source models to robustly conclude that the system is dominated by extended emission within 50 au. 
We report detections of two previously observed asymmetric rings at $\sim$17 au and $\sim$45 au. The peak brightness location of each model ring is consistent with the previous observations. We also, for the first time with imaging, robustly recover an elliptical Gaussian inner disk, previously inferred via SED fitting. This inner disk has a FWHM of ~5 au and a similar inclination and orientation as the outer rings. Finally, we recover no clear evidence for candidate planets. By modelling the extended disk emission, we are able to place a lower limit on the near infrared companion contrast of at least 1000. 
\end{abstract}




\section{Introduction} 
\label{sec:intro}

Protoplanetary disks are the expected sites of planet formation and are thus prime targets for young planet searches. Many protoplanetary disks, especially transition disks, have observable substructure in the millimeter (mm) dust continuum and molecular lines \citep{2018ApJ...869L..42H,2011ApJ...732...42A,2016A&A...585A..58V} that is similar to that expected from planet-disk interaction theory \citep{2012A&A...545A..81P}. Observations of these structured systems at high angular resolution in the near infrared allow us additional characterization of the inner dust disk and potentially the discovery of planets in formation.
{Beyond just the discovery of forming planets, it is expected that these data will increase our understanding as to the accretion of material onto the planet in connection with the transport of material through the disk gap \citep[e.g.][]{2006ApJ...641..526L}}. 

LkCa\,15 is a nearby (160 pc), young, T Tauri star with spectral type K2, mass 1.32 M$_\odot$, and luminosity 1.3 L$_\odot$ \citep{Francis_2020} surrounded by a protoplanetary disk with a central dust depleted cavity and three wide-orbit, narrow dust rings observed in the mm at $\sim$47, $\sim$69, and $\sim$100 au \citep{2020A&A...639A.121F}. LkCa\,15 was classified as a prototype for the pre-transition disk class of objects by \citet{Espaillat_2007} due to the presence of a significant near-infrared excess over the stellar photosphere in the spectral energy distribution, suggesting a compact and optically thick inner disk. An unresolved inner disk may also be present at mm wavelengths, as deep ALMA observations reveal significant but low-level emission within the cavity located interior to the  $\sim$47 au ring \citep{2020A&A...639A.121F}.

Previous studies of LkCa\,15 utilizing near infrared sparse aperture masking interferometry (SAM) claimed to have observed planets in the process of formation embedded in its disk. \cite{Kraus_2011} reported a multi-epoch detection of a single blue compact source (deprojected orbital
radius $\sim$16–20 au) surrounded by redder material using SAM on Keck II. \cite{Sallum_2015} reported the detection of three point sources (LkCa\,15 b,c,d with deprojected orbital
radii between $\sim$15–19 au) using SAM on the Large Binocular Telescope. They also detected LkCa\,15b in the accretion-tracing H$\alpha$ emission using SDI with the Magellan Adaptive Optics System.  More recent direct imaging work, however, using SPHERE ZIMPOL and IRDIS \citep{2015ApJ...808L..41T,2016ApJ...828L..17T} and Subaru/SCExAO coupled with CHARIS, and Keck/NIRC2 data have disputed some of these results \citep{Currie_2019}, by showing that the SAM detections of LkCa\,15 b,c,d can likely be attributed to disk emission from a ring with an inner radius of $\sim$20 au.

Sparse aperture masking interferometry \citep[SAM;][]{1986Natur.320..595B,1987Natur.328..694H,2000PASP..112..555T,2015ApJ...798...68G} is a high resolution optical interferometric observing technique that uses a mask (at the pupil plane) to transform the full telescope aperture into a set of smaller sub-apertures. Each sub-aperture pair acts as a two station interferometer, sampling a discrete component of the Fourier transform of a target's brightness distribution. SAM can resolve objects with tighter separations than can be resolved with a filled aperture \citep{2006ApJ...650L.131L,Sallum_2019}, making it a good technique for observing planets at close separations from their stars. Due to the sparse nature of SAM data (tens of data points per exposure), accurate image reconstruction is difficult and thus inner disk signals can be confused with planet detections \citep{2013ApJ...762L..12C,2013ApJ...768...80K}.

The van Cittert–Zernike theorem \citep{VANCITTERT1934201, ZERNIKE1938785} states that the complex visibility to go from the image plane to the pupil plane is given by the two dimensional Fourier transform
\begin{equation}
    V(u,v) = \int I(x,y) \exp (-i2\pi(ux+vy)) dydx.
\end{equation}
where $u$ and $v$ are the coordinates in the Fourier domain (measured in units of the observing wavelength), $x$ and $y$ are the coordinates in the image plane, and $I(x,y)$ is the target brightness distribution. It is not currently possible to obtain well-calibrated phases on long baselines from ground based observations due to calibration challenges. Thus, more robust quantities are often derived from the individual complex visibility measurements, with an accompanying slight loss in information content. The first robust SAM observable is the squared visibility, or the modulus squared of the complex visibility. The second observable is the closure phase. The closure phase is a source-dependent quantity made by summing phases around triangles formed by holes in the mask. The closure phase is useful because the phase errors introduced by the telescope can be removed \citep{2007NewAR..51..604M}; phase shifts produced by noise sources, such as the atmosphere or electronics, will cause equal but opposite phase shifts for connected baselines in each closure triangle. 


As previously noted, a difficulty of working with SAM observations is that the data can be sparse, only consisting of a limited number of squared visibilities and closure phase measurements. This makes reconstructing an image of arbitrary complexity difficult, if not impossible. The lack of an easily obtainable, reliable image reconstruction becomes especially challenging when prior knowledge about a target is limited.  Given sufficient prior knowledge about the types of objects being observed, however, a  powerful approach for analyzing sparse interferometric data can be to use geometrical models. Geometrical models are constructed using analytic descriptions to model the major components in the scene using only a limited number of parameters.

Here we present an analysis of new near infrared SAM observations obtained by VLT/SPHERE-IRDIS \citep{2016SPIE.9907E..2TC} of the transition disk LkCa\,15. We use Bayesian model fitting techniques to determine the structure of the inner tens of au around LkCa\,15, testing the limits of using these high resolution techniques for observing the inner regions of protoplanetary disks. We also discuss the application of these methods to observing planets within transition disks. The structure of the paper is as follows: \S\,\ref{sec:obs} outlines how the data were obtained and processed; \S\,\ref{sec:methods} details the techniques used for image reconstruction, including the geometrical models that were fit; \S\,\ref{sec:analysis} outlines how the data were analysed using these techniques and the results of applying these techniques; \S\,\ref{sec:discussion} discusses the image reconstructions and performs a companion analysis, where point source models are considered in combination with the more complex geometrical models. We close with a summary in \S\,\ref{sec:conclusions}.

\section{Observations}
\label{sec:obs}

\begin{figure}[ht]
    \includegraphics[width=0.95\columnwidth]{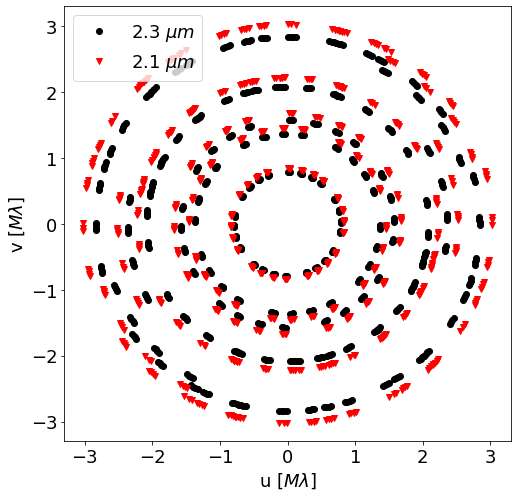}
    \caption{LkCa\,15 VLT/SPHERE-IRDIS uv-plane coverage. The largest and smallest baselines, $B$, correspond to spatial scales ($\lambda /B$) $\sim$70-260 mas (10-40 au). }
    \label{uv_cov}
\end{figure}

LkCa\,15 and calibrator star HD\,284581 were observed at 2.11 (2.1) and 2.25 (2.3)\,$\mu$m  \citep[K12 dual-band imaging filters,][]{2010MNRAS.407...71V} on January 4, 2018, using the 7 hole (21 baseline) VLT/SPHERE-IRDIS \citep{2019A&A...631A.155B, 2008SPIE.7014E..3LD} SAM mode \citep{2011Msngr.146...18L,2010SPIE.7735E..1OT,2016SPIE.9907E..2TC}, over the course of about two hours. {The spatial scales covered by our observations (see Figure \ref{uv_cov}), defined by the longest and shortest baselines, are $\sim$70-260 mas or 10-40 au at the distance of LkCa\,15. Note that the standard diffraction limit of a telescope with an aperture the size of the longest baseline would be 83 mas. Due to the interferometric nature of our data, we anticipate sensitivity to scales as small as $\sim$35 mas or $\sim$5 au.} The LkCa\,15 data set comprises four exposures, with each consisting of four integrations. Three exposures with the same number of integrations and frames per exposure were acquired for the calibrator HD\,284581. 

The IRDIS data were reduced using the SPHERE Data Reduction and Handling (DRH) automated pipeline \citep{2008SPIE.7019E..39P}. We applied basic corrections for bad pixels, dark current, and flat field, as well as distortion correction. Absolute orientation and pixel scales of the images were calculated using the parallactic angle and the absolute calibration provided by the SPHERE consortium. The current respective estimates of the pixel scale and true north for IRDIS are 12.255$\pm$0.009 mas/pixel and TN=\mbox{-1.75$\pm$0.08$^{\circ}$} \citep{2016SPIE.9908E..34M}. 
Our observations were obtained using the pupil-stabilized mode. The absolute orientation (north-up, east-left) is therefore retrieved within the DRH using both TN and pupil offset (-135.99$\pm$0.11$^{\circ}$).

AMICAL \citep{2020SPIE11446E..11S} was used to extract the closure phase, squared visibility, and uncertainties from each integration, as well as to calibrate the data. The Fourier extracting method applied within AMICAL required additional reduction steps to allow an optimal extraction. The processed data (post-DRH) were background subtracted, cropped, and cleaned of residual bad pixels and cosmic rays. The cleaned cubes were then windowed with a super-Gaussian function of the form, $\exp({-ar^4})$, before closure phases and visibilities were extracted from the Fourier transforms of the images. Each complex visibility (over 21 baselines) is computed over a sub-sample of pixels using a weighted average. These pixel weights are retrieved from the absolute positions of the 7 apertures (in meters from the pupil plane) and the pixel size of IRDIS. The appropriate combination of Fourier transform for each aperture  pair takes into account the fraction of pixel not centered on the real pixel position and increases the accuracy of the method\footnote{See \citep{2020SPIE11446E..11S} for details.}. A $\sigma$-clipping selection was applied to reject eventual bad frames due to bad seeing conditions or AO instability. Calibration of the closure phases (visibilities) was performed on each wavelength individually by subtracting (dividing) a weighted sum of the corresponding measurements taken on the nearest in time integration of the calibrator star. The extracted observables were finally saved as standard interferometric file format oifits \citep{2017A&A...597A...8D}, taking into account the astrometric orientation obtained from the DRH (i.e., rotation of the u, v coordinates). The observables extracted from one of the 2.1 $\mu$m integrations were discarded due to a large variance from the mean compared to the other extracted data points. 

{In order to optimally utilize the available data, in our analysis we assume that the colour variations between 2.1 and 2.3 $\mu$m across the components of LkCa\,15 are not significant and therefore that the benefit of increasing uv-plane coverage (see Figure \ref{uv_cov}) outweighs the additional reconstruction uncertainty obtained by considering the two wavelengths together during image reconstruction.}

\section{SAM ANALYSIS METHODS}
\label{sec:methods}

Sparse aperture masking interferometry is a technique well suited for observing multiple point sources, such as a binary star system or a star orbited by a bright planet. Such sources have a distinct signature in the Fourier domain that is simple enough to often only require tens of samples to robustly observe. For two point sources (with one at the phase center) the complex visibilities 
{at the observed coordinates in the $u$, $v$ plane} are given by the four parameter formula
\begin{equation}
    V_{binary}(u,v) = \alpha + \beta \exp(-i 2 \pi (ux_0 + vy_0)),
    \label{binary_vis_eqn}
\end{equation}
where $\alpha$ is the central star brightness, $\beta$ is the off-center source brightness, and $x_0$ and $y_0$ are the coordinates of the off-center point source in the image domain.

Interpreting SAM data can be significantly more challenging when complex structures beyond point sources are being observed. Image reconstruction from sparse data is an ill-posed, under-constrained inversion problem, that can be addressed with the use of priors. One method for using strong prior knowledge about a target is to utilize image domain geometrical models. Instead of having to determine complex analytic Fourier transforms for any arbitrary geometry, the fast Fourier transform can be used. Here, the complex visibilities of geometrical models are calculated using a fast-Fourier-transform (FFT) and bi-linear interpolation similar to what is done in GALARIO \citep{tazzari}. This technique is implemented using Tensorflow-GPU \citep{tensorflow2015-whitepaper} for a significant speed-up compared with other available methods.

\subsection{Geometrical Disk Models}
\label{sec:geo}

In this work we primarily use a simple geometrical model to describe a disk component with a lopsided brightness distribution: the polar Gaussian ring (sometimes referred to in this paper as an arc). This model has been frequently used for model fitting to mm interferometer observations of disks and rings with azimuthal asymmetries \citep[e.g.][]{2015A&A...584A..16P,2021AJ....161..264H}, but does not capture the full three dimensional geometry of a thick disk \citep{dong15gap}. To account for the dominant emission from the central star, a bright central point source is also included analytically in the Fourier domain. 

The intensity distribution of the polar Gaussian ring, in the $r$, $\theta$ plane of the disk, is given by a five-parameter formula
\begin{equation}
\label{pg}
    I_{GR}(r,\theta) = I_0 \exp\left(-\frac{(r-r_0)^2}{2\sigma_r^2} - \frac{(\theta-\theta_0)^2}{2\sigma_\theta^2}\right),
\end{equation}
where $r_0$ is the central radius of the ring, $\theta_0$ is the position angle (defined East of North, where East is positive RA and North is positive DEC) of the bright Gaussian peak, $\sigma_r$ defines the radial width of the ring, $\sigma_\theta$ defines the azimuthal width of the arc, and $I_0$ is the peak brightness in the ring. Requiring the origin of the ring to be fixed at the central star allows for only two additional fitting parameters: the observed disk inclination, $i$, and position angle, PA. Under the assumption that the brightness asymmetry is due to scattering from the inclined disk, however, we equate the position angle with the direction of the peak brightness in the ring such that PA$= \theta_0$. Finally, we note that the scattered light emission from a thick disk may not follow exactly a circular arc in the plane of the disk {\citep[e.g.,][]{Dong__2015}}. Thus, for some geometrical fits we allow the origin of the ring to deviate from the central star position. 

To account for the possibility of multiple rings in LkCa\,15, additional polar Gaussian ring components can be included in the fit. Furthermore, it is likely that a (marginally) resolved inner disk exists around the central star \citep{Espaillat_2007, 2010A&A...512A..11M,2018A&A...620A.195A} and thus in some cases we also allow for a central, inclined Gaussian ($\propto e^{-(x^2+(y/cos(i))^2)/2}$) disk component to model this emission. The Gaussian disk component is allowed to have its own inclination and position angle, as well as a small offset from the central star.

\subsection{Bayesian Model Fitting}
\label{sec:bayesian}

We use Bayesian model fitting of the LkCa\,15 visibilities which allows explicit priors to be included and model comparison to be conducted. We assume our set of measurements are independent Gaussian random variables, so the likelihood, or the probability of obtaining a set of $n$ measurements given a model $M$ with parameters $\theta$, is given by
\begin{equation}
    \label{like}
    P(D|\theta,M) = \frac{1}{\prod_{i}^{n}\sqrt{2\pi \sigma_i^2}}\exp\left(\sum_{i}^{n} \frac{-(y_i - y(x_i))^2}{2\sigma_i^2}\right),
\end{equation}
where $y_i$ is an individual measurement with measured standard deviation $\sigma_i$ and $y(x_i)$ is the prediction generated by $M$. For the entirety of our analysis we assume flat priors on all model parameters.

To fit models to the data by sampling the posterior probability, dynamic nested sampling \citep{dyn2018} with \texttt{dynesty} \citep{Speagle_2020} is used. Nested sampling \citep{2004AIPC..735..395S,10.1214/06-BA127} is a powerful method because, unlike with MCMC based methods, an initial guess for model parameters is not used, so the entire allowed parameter space is efficiently mapped in an un-biased manner. Nested sampling also directly calculates the (model) evidence, or marginal likelihood, which can be used for model comparison. The evidence, $P(D|M)$, is defined as the integral over the entire set of model parameters, $\Omega_\theta$, of the likelihood, $P(D|\theta,M)$, multiplied by the prior, $P(\theta|M)$,

\begin{equation}
\label{evidence}
    P(D|M) = \int_{\Omega_\theta} P(D|\theta,M) P(\theta|M) d\theta.
\end{equation}
Our derived model parameters are chosen to be the median of the marginalized posterior distributions and our uncertainties are the 16th and 84th percentiles of the posteriors. The images produced by each model are 256$\times$256 pixels with a pixel size of 4.7 mas and a central star brightness fixed at log$_{10}\,I_{s} = 12$, in the normalized units used throughout this analysis. We use dynamic nested sampling with flat priors, 1500 initial live points, and batches of 500 live points added until \texttt{dynesty}'s default convergence criteria are satisfied.

\subsection{Direct Image Reconstruction}
\label{sec:imre}

Many effective, well tested algorithms exist for directly reconstructing images from sparse interferometric data [e.g. \texttt{SQUEEZE} \citep{squeeze} and \texttt{MiRA} \citep{2008SPIE.7013E..1IT}]. 
However, for this work a simple stochastic gradient descent based algorithm was written in Python using Tensorflow \citep{tensorflow2015-whitepaper}. This algorithm takes advantage of Tensorflow's GPU compatibility, and uses auto-differentiation \citep{10.2307/2689402} to calculate derivatives. The code consists of a single `locally-connected' layer \citep{gregor2010emergence} that takes an initial `prior' image as input and multiplies each pixel value by a trainable parameter that is initialized to one. For our analysis, compactness and total variation regularizers were used \citep[see,][]{refId0}. The compactness hyper-parameter value was chosen by inspecting the gradients to determine the optimal value to concentrate the flux within the region of highest sensitivity for our data (R $\lesssim$ 300 mas). The total variation hyper-parameter was found using the L-curve method \citep{doi:10.1137/1034115}.

The algorithm uses the Adam optimizer \citep{kingma2017adam} with a learning rate of 0.1 to reconstruct an image by minimizing the reduced $\chi^2$ of the output image closure phase and squared visibility measurements plus the regularization term.

\section{Analysis of the VLT/SPHERE-IRDIS Data} 
\label{sec:analysis}

\subsection{Geometrical Model Reconstruction}
\label{sec:gmr}

\begin{figure}[ht]
    \includegraphics[width=0.89\columnwidth]{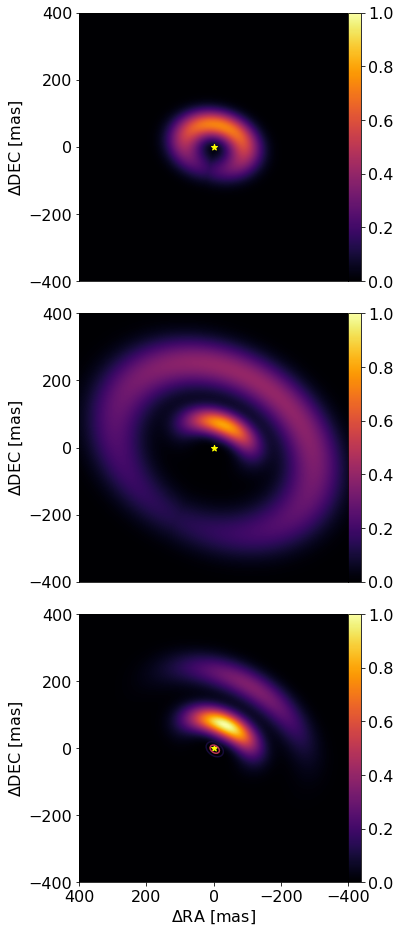}
    \caption{Best fit geometrical models based on \texttt{dynesty} fits to 2.1 and 2.3 $\mu m$ VLT/SPHERE-IRDIS closure phase and squared visibility data of LkCa\,15 (images are shown on the same fixed intensity scale with the normalization such that the peak polar Gaussian ring component brightness in the bottom panel is one).
    \textit{Top}: Single polar Gaussian ring model (1PG); \textit{Middle}: Two polar Gaussian rings model (2PG);
    \textit{Bottom}: Two polar Gaussian rings plus central, inclined, Gaussian model (2PG+IG), with the inner Gaussian component shown by its half maximum and one tenth maximum contours (same colour scale but normalized by its peak brightness).} 
    \label{6_panel}
\end{figure}

Previous near infrared observations of LkCa\,15 have attributed the observed emission to scattered light from a disk surface \citep[e.g.][]{2015ApJ...808L..41T,2016ApJ...828L..17T,Currie_2019}, motivating our use of the arc-like polar Gaussian ring geometrical model (\S \ref{sec:geo}). This model, and each subsequent multi-component model, were fit to the combined 2.1 and 2.3\,$\mu$m data for LkCa\,15 taken by VLT/SPHERE-IRDIS by comparing calculated model squared visibilities and closure phases to the data. This analysis was performed with dynamic nested sampling using \texttt{dynesty} \citep{Speagle_2020} to find the most likely model parameters, to approximate the posterior probability distributions, and to calculate the Bayesian evidence for model comparison.
 
To help avoid issues with potential degeneracy between the radius and inclination of our ring models, the inclination of LkCa\,15 is constrained to be between 40 and 60 degrees, as LkCa\,15 is known to have a moderately inclined disk on the sky \citep[e.g.][]{2014A&A...566A..51T,10.1093/pasj/psv133}. All other parameters are constrained only to be physically realistic. In addition to the geometrical model parameters, two normalization parameters are included in order to scale the calculated squared visibilities and to correct for the fact that we fit a single model with data obtained at two wavelengths.
 
As a first test, a single polar Gaussian ring (1PG) was fit to the data. The best fit model (top panel in Figure \ref{6_panel}) from the dynamic nested sampling, {has a similar geometry and orientation to the near infrared LkCa\,15 disk emission features \citep{Currie_2019} previously observed via coronagraphy,} and probes angular scales well sampled by the SAM observations, $\sim$100 mas. The single polar Gaussian ring model has a log evidence, ln$(P(D|M)) = -884.2 \pm 0.2$, indicative of a rather poor fit when compared to other tested models (see below) and suggesting that the  model is overly simple to accurately reproduce the LkCa\,15 observations. 
 
An outer disk, beyond the arc seen in the single polar Gaussian ring model, has been observed previously at similar wavelengths to our data \citep[e.g.][]{2014A&A...566A..51T,10.1093/pasj/psv133,Currie_2019}. The bright inner edge of this outer disk component resides at an angular scale which the present SAM observations can probe and therefore should contribute a non-negligible amount to our visibility measurements. We included this outer disk using a second polar Gaussian ring (2PG; middle panel in Figure \ref{6_panel}), where to avoid a degeneracy between the rings the de-projected central radius of the outer ring is constrained to be larger than 200 mas. Additionally, the position angle and inclination of the outer ring are constrained to be offset from the inner ring by at most $\pm$ 10 and 5 degrees, respectively. For this model, the parameters estimated using \texttt{dynesty} give an inner arc comparable to what is seen in the single polar Gaussian ring image, and a reasonably well constrained faint outer arc, similar to what is seen by \citet{2014A&A...566A..51T}, \citet{10.1093/pasj/psv133}, and \citet{Currie_2019} (see also Section \ref{sec:compare}). The two component polar Gaussian ring model is significantly more consistent with the VLT/SPHERE-IRDIS visibility measurements, having ln$(P(D|M)) = -577.7 \pm 0.2$.
 
The final component added to the modeling was a marginally resolved central, inclined (elliptical) Gaussian disk, to account for emission on scales less than {$\sim$}50 mas ($< 8$\,au). The VLT/SPHERE-IRDIS observations provide a unique opportunity to uncover structure at this angular scale.  There exists independent observational evidence for an inner disk ($\sim$1\,au) in LkCa\,15 from spectroscopic observations \citep{2018A&A...620A.195A} and one has been hypothesized in spectral energy distribution (SED) studies \citep[e.g.][]{Espaillat_2007,2010A&A...512A..11M,2010ApJ...717..441E}. A sub-au disk component also was used in the LkCa\,15 radiative transfer modelling by both \cite{2014A&A...566A..51T} and \cite{Currie_2019}. The best fit model image for the two polar Gaussian rings plus a central, inclined, Gaussian disk (2PG+IG) is shown in the bottom panel of Figure \ref{6_panel}. By including the central disk an additional large improvement in the quality of the fit is obtained, with an increase in the log evidence to ln$(P(D|M)) = -227.7 \pm 0.3$.  The model parameters derived for this three component model by the dynamic nested sampling are shown in Table \ref{2pg_par_bf}. 

For completeness, the 2PG+IG model was also tested while allowing for a single offset from the origin for both of the polar Gaussian components. A nearly identical solution to the non-offset model was found within the uncertainties, with only a small, insignificant, spatial offset required ($\Delta$\,DEC = $4^{+9}_{-13}$\,mas and $\Delta$\,RA = $-4^{+6}_{-4}$\,mas), indicating that there is no requirement for these extra parameters. We also explored a small-scale stellar binary model as an alternative to an inner Gaussian disk but obtained a much poorer fit to the data, ln$(P(D|M)) = -386.6 \pm 0.3$.

\begin{deluxetable*}{ccccccc}
    \label{2pg_par_bf}
    \tablecaption{
    Two polar Gaussian rings plus central, inclined, Gaussian disk model (2PG+IG) parameters estimated from dynamic nested sampling marginalized distributions. Upper and lower uncertainties are the 84th and 16th percentiles, respectively. The central point source intensity was added as a constant in the Fourier domain and was fixed at log$_{10}\,I_s$ = 12.}
    
    \tablewidth{0pt}
    
    \tablehead{{Inner Arc}&log$_{10} I_0$&$r_0$ & $i$ &PA & FWHM$_r$&FWHM$_{\theta}$\\
    \nocolhead{name} &\colhead{[pix$^{-1}$]} &\colhead{[mas]}& \colhead{[deg]} &\colhead{[deg]}  & \colhead{[mas]}&\colhead{[deg]} }
    \startdata
     &$7.78^{+0.02}_{-0.02}$& $107^{+3}_{-3}$ & $44^{+2}_{-2}$ & $-24^{+1}_{-1}$& $87^{+6}_{-6}$ & $98^{+6}_{-6}$ \\
     \hline
    {Outer Arc}&log$_{10} I_0$&$r_0$ & $\Delta i$&$\Delta$PA& FWHM$_{r}$&FWHM$_{\theta}$ 
    \\
  \nocolhead{name}&\colhead{[pix$^{-1}$]} &\colhead{[mas]}& \colhead{[deg]} &\colhead{[deg]}  & \colhead{[mas]}&\colhead{[deg]} 
    \\ \hline
    &$7.34^{+0.04}_{-0.05}$&$285^{+11}_{-9}$  & $-2.2^{+2.4}_{-1.8}$ & $-5.9^{+1.2}_{-1.4}$ & $104^{+9}_{-9}$ & $70^{+8}_{-8}$ \\
      \hline    Central Disk&log$_{10} I_G$&FWHM& $i$&PA&  $r_0$$^a$& $\phi_0$$^a$
     \\
   \nocolhead{name}&\colhead{[pix$^{-1}$]} &\colhead{[mas]}& \colhead{[deg]} &\colhead{[deg]}  &\colhead{[mas]} & \colhead{[deg]} 
    \\ \hline
    &$9.45^{+0.24}_{-0.22}$&  $32^{+5}_{-5}$ &  $46^{+2}_{-2}$& $-34^{+3}_{-3}$& $1.2^{+0.2}_{-0.2}$ & $239^{+20}_{-19}$ 
    \enddata
    \footnotesize{$^a$ The parameters $r_0$ and $\phi_0$ define the offset of the central disk in polar coordinates where $\phi$ is defined in the same manner as the position angle.}
\end{deluxetable*}


\subsection{Direct Image Reconstruction}

The direct image reconstruction technique described in \S\,\ref{sec:imre} was used to further investigate the robustness of the geometrical model fits. Details of the results are provided in Appendix  \ref{app:dir} and here we present an overview of the main results.


From starting with an axisymmetric smooth distribution of emission, after 250 iterations the direct reconstruction technique recovers a feature very similar to the inner  arc found by all our geometrical models, 1PG, 2PG, and 2PG+IG. In addition, with this direct reconstruction,  marginally extended bright emission remains near the central star, similar to the central Gaussian disk component found in the 2PG+IG geometrical model. Finally, the outer arc found by the 2PG and 2PG+IG geometrical model fits, at angular distances poorly measured by the SAM observations, is not recovered. 

To test the robustness of the direct image result, we followed a similar approach to that used by \citet{sanchez2021} and directly reconstructed an image using simulated data from the best fit 2PG+IG model instead of the observations themselves. For this test we obtain a very similar image to that discussed in the previous paragraph. Furthermore, an image was directly reconstructed from simulated data of a multiple point source model, consisting of LkCa\,15 and three candidate planets.  The reconstructed image clearly retains each of the three distinct point sources. In combination, the direct image reconstructions, although utilizing a significantly under-sampled measurement set, yield results similar to our expectations and thus provide additional confidence that the geometrical models presented here represent robust solutions.

\section{Discussion}
\label{sec:discussion}

\subsection{Comparing Disk Image Reconstructions}
\label{sec:compare}

The best fit properties for the three component geometrical model (2PG+IG) including an inner and an outer polar Gaussian ring plus a central, inclined, Gaussian disk are provided in Table \ref{2pg_par_bf}. The inner Gaussian (central disk) component in our model has a FWHM $\sim$5 au, the radii of the inner and outer polar Gaussian rings in our model are $\sim$17 au and $\sim$45 au, respectively, with an inclination of $\sim$45 degrees.  On scales of the inner and outer rings, our model is mostly consistent with what has been previously quantified at similar wavelengths using direct {or coronagraphic} imaging by \citet{2014A&A...566A..51T,2015ApJ...808L..41T,2016ApJ...828L..17T}, \citet{10.1093/pasj/psv133} and \citet{Currie_2019}. 
The compact central Gaussian disk of our three component model is near the smallest angular scales that we expect to be able to observe with our SAM measurements. This central disk corresponds to the innermost component in the models by \citet{2014A&A...566A..51T,2015ApJ...808L..41T} and \citet{Currie_2019}, which has been inferred from SED fitting \citep[e.g.][]{2010A&A...512A..11M}. Our inner ring corresponds to the inner ring observed by \citet{2015ApJ...808L..41T,2016ApJ...828L..17T}, \citet{10.1093/pasj/psv133} and \citet{Currie_2019}. This component resides at an ideal angular scale given our SAM uv-plane coverage ($\lambda/B \sim$ 70-260 mas). Our outer ring resides at the same location as the original ring that was found \citep{2007ApJ...670L.135E} and which has been identified by all of the above studies. This ring is near the outer edge of the spatial scales to which our SAM observations are sensitive.  

\begin{figure*}
   \centering
    \includegraphics[width=0.95\textwidth]{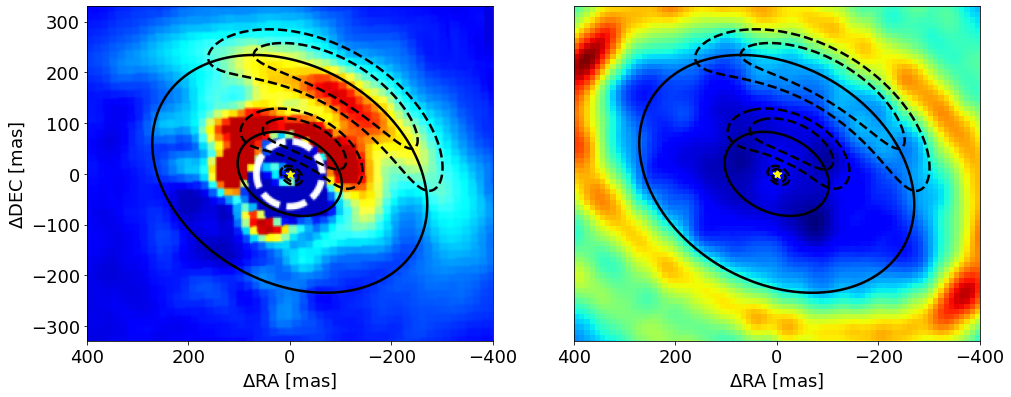}
    \caption{
    \textit{Left}: Ring locations (solid lines) and half-maximum/quarter-maximum brightness contours (dashed lines) for the best fit two polar Gaussian rings plus central, inclined, Gaussian disk model (2PG+IG) plotted over the K-band coronagraphic image of LkCa\,15 by \cite{Currie_2019}. The white dashed circle denotes a radius of 60 mas, {approximately the inner working angle of the CHARIS instrument \citep{10.1117/12.2025081}}. 
    \textit{Right}: The same model plotted over the mm wavelength ALMA image of LkCa\,15 by \citet{2020A&A...639A.121F}.}
    \label{fig:cont_over_curr}
\end{figure*}

In Figure \ref{fig:cont_over_curr} we plot the 50\% brightness contours for each component of our best fit model over the K-band {coronagraphic} image by \citet{Currie_2019} to visually emphasize the agreement discussed above. The inner arc very closely matches the radial location and extent seen in the K-band {coronagraphic} image. The outer arc, however, does not match exactly the emission seen in the  K-band image. Although the position angle of the peak brightness is in good agreement, the radial extent of the model is shifted somewhat outward. This small discrepancy is likely attributable to our decreased sensitivity at larger separations and (relatively) lower contrast. Additionally, the polar Gaussian model used here is extremely simple and therefore does not take into account expected complexity in the scattered light emission structure \citep{Dong__2015}.  
In Figure \ref{fig:cont_over_curr} we also plot our model contours over an ALMA image of LkCa\,15 by \citet{2020A&A...639A.121F}, which demonstrates that the rings we find are interior to the rings seen with ALMA.

\begin{deluxetable}{cccc}
    \label{comp_par}
    \tablecaption{
        Derived disk parameters from 
        \citet{10.1093/pasj/psv133} and \citet{Currie_2019}. The ellipse fits contain a sometimes considerable offset from the star not shown here.}
    
    \tablewidth{0pt}
    
    \tablehead{Outer Ring&$r_0$ & $i$ &PA \\
    \nocolhead{name} &\colhead{[au]}& \colhead{[deg]} &\colhead{[deg]}  }
    \startdata
     Oh (2016)$^a$ & 59.0 $\pm$ 1.4&44 $\pm$ 1& -31 $\pm$ 2\\
     Currie (2019)$^b$ & 55  &50& -30 \\
     This work & $45^{+2}_{-1}$ & 43 $\pm$ 2 & -30 $\pm$ 1 \\
     \hline
    Inner Ring&$r_0$ & $i$&PA\\
  \nocolhead{name}&\colhead{[au]}& \colhead{[deg]}&\colhead{[deg]} 
    \\ \hline
    Oh (2016)$^a$ & 29.8 $\pm$ 2.0 &44 $\pm$ 2 & -18 $\pm$ 2 \\
    Currie (2019)$^b$ & 20 &51.5 & -30 \\
    This work & 17.0 $\pm$ 0.5 & 44 $\pm$ 2 & -24 $\pm$ 1 \\
    \enddata
\footnotesize{
$^a$ Ellipse fit to middle of ring. 

$^b$ Radiative transfer model (radii are inner edge of each disk component).}
\end{deluxetable}

\begin{figure*}
    \centering
    \includegraphics[width=0.85\textwidth]{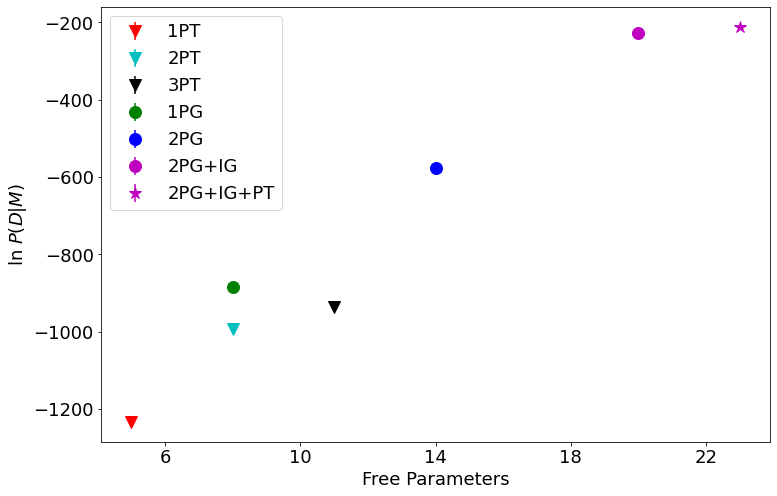}
    \caption{The evidence ($\ln (P(D|M))$)  for each model discussed in this paper as a function of the number of free parameters for the model. Recall that the evidence takes into account the number of free parameters and thus can be fairly compared across these models. PT = point source companion (with 1 to 3 companions), PG = polar Gaussian ring (with 1 or 2 independent rings), IG = inclined Gaussian.
    }
    \label{fig:all_like}
\end{figure*}

Our best fit model was built up iteratively, starting from a single arc, with the evidence improving with each added component, as shown in Figure \ref{fig:all_like}. {These log evidence values can be quantitatively compared using the Bayes factor \citep{doi:10.1080/01621459.1995.10476572,bayesevid2011}, the ratio of evidences, which says that there is strong evidence for one model in favour of another when the difference between the log evidences is greater than 5 \citep{2008ConPh..49...71T}}. The large increases seen in the evidence with each included model component provides a strong indication that such components contribute significantly to the disk signal present in our visibility measurements. 
{This is an important result because it shows that the SAM observations have leverage over an extended range in angular scales, since both the outer arc and compact central disk are near the edge of the scales probed by VLT/SPHERE-IRDIS SAM. Furthermore, this result reveals the need to model each of these components before hunting for faint forming planets.} 

Our procedure uses a minimum number of free parameters, and thus the derived geometrical values are not necessarily a true reflection of the inherently 3-dimensional structure of the LkCa\,15 disk components. This consideration needs to be taken into account when comparing our derived parameters with previous determinations from direct imaging studies (Table \ref{comp_par}). Due to the (relative) simplicity of our models, it is satisfying to see the strong agreement in position angle and inclination while not surprising that the radius measures have somewhat larger deviations. {Unlike the fits in the previous papers, our rings do not have appreciable offsets from the star, and we have only included a single polar Gaussian for each ring, which is unable to capture the full azimuthal extent of the disk.}

To derive more accurate physical properties for the LkCa\,15 disk it will be necessary to consider its three-dimensional structure. The best approach to accomplish this task would be to use radiative transfer models as the input parameters to the visibility measurement fitting. The difficulty with this approach is the computational cost of radiative transfer models. In this work we were able to evaluate millions of models in minutes, whereas individual three-dimensional radiative transfer simulations, in the mostly optically thick regime our observations occupy, can take on the order of minutes each. This motivates the development of accelerated, or approximate, radiative transfer modelling techniques.

\begin{figure*}
    \includegraphics[width=0.95\textwidth]{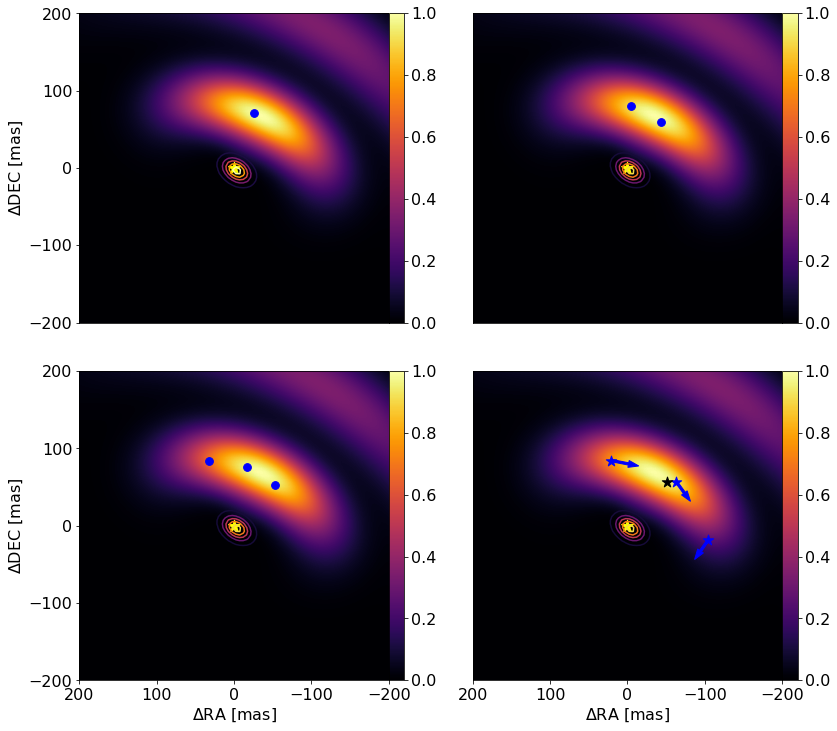}
    \caption{Best fit locations for one (1PT, top left), two (2PT, top right), and three (3PT, bottom left) companions when modeling only for point sources, plotted over the best fit 2PG + IG model, from \texttt{dynesty} fit of 2.1 and 2.3 $\mu m$ closure phase and squared visibility data of LkCa\,15. The bottom right panel shows the 2014 L' detections of LkCa\,15 b,c,d from \cite{Sallum_2015} as blue stars, the 2009 L' planet detection from \cite{Kraus_2011} as a black star, and the approximate predicted LkCa\,15 b,c,d locations at the time of our observations (blue arrows).}
    \label{pt_sources}
\end{figure*}

\subsection{Importance of Including Extended Disk Components when Searching for Planets}
\label{sec:companion}
\subsubsection{Planets versus Rings}

To test whether high contrast, small separation, extended structure can be robustly distinguished from point like emission the same reconstruction method as utilized to fit for rings (\S\,\ref{sec:gmr}) was applied to models consisting of multiple point sources. We therefore fit for one (1PT), two (2PT), and three (3PT) planets orbiting LkCa\,15, under the assumption of no extended disk emission. In Figure \ref{fig:all_like} we present the evidence for each of our multiple point source models while in Table \ref{binaries} we tabulate the best fit parameters. Although the evidence increases with each additional included point source, even the eleven parameter, three companion, model (3PT) has poorer evidence compared to the eight parameter single polar Gaussian ring model (1PG) described above. This clearly demonstrates that the largest contribution to the  visibility measurements is a feature consistent with smooth extended arc-like emission, rather than point-like sources. 

We also investigated where in the image plane the best fit point sources are located when no extended structure is included in the model. Figure \ref{pt_sources} shows the locations of the point source fits plotted over the (zoomed in) three component geometrical disk model (2PG+IG). As expected, the point source locations trace the peak brightness of the inner disk arc. Similarly, the bottom right panel in Figure \ref{pt_sources} shows the claimed planet detections from \cite{Kraus_2011} (black star) and \citep{Sallum_2015} (blue stars). Our point source fits are consistent with the previously determined candidate planet positions, with the exception that we do not find any point sources coincident with LkCa\,15b. This is not surprising as there is only one 
single epoch H$_\alpha$ detection of LkCa\,15 b \citep{Sallum_2015}, and more recent work \citep{2019hsax.conf..359M} has argued that LkCa\,15's $H_{\alpha}$ emission is more consistent with a disk than a planet. {We also recover contrasts of a few hundred (see Table \ref{comp}), which are mostly consistent with the contrasts reported by \citet{Sallum_2015}}.  We emphasize, however, that our point source model fits are significantly poorer than the extended structure model fits and included here as a demonstration of the potential pitfalls in interpreting sparse data sets using models incompatible with the underlying source structure.

Along with the goodness of fit criterion, the time-domain can be used to separate planets from disk emission. Planets should orbit at Keplerian rates around the central star. In contrast, the lopsided brightness observed from an inclined disk should remain fixed in location. To illustrate this difference, blue arrows in the bottom right panel of Figure \ref{pt_sources} denote the expected orbital locations of the previously proposed planets at the time of our observations, following the prescription in \citet{Currie_2019}. As the Keplerian orbits are long, significantly more time between observations would be required to capture observable changes in the viewing geometry. For systems with disks, however, recovering Keplerian motion is likely to be a key indicator of robust planet detection.


\begin{deluxetable*}{cccccccccc}

  \tablecaption{\label{binaries}
  Companion parameters estimated from dynamic nested sampling marginalized likelihood distributions. Upper and lower uncertainties are the 84th and 16th percentiles, respectively.}

    \label{comp}
    \tablehead{Model&$r_0$ [mas]&PA$_0$ [deg.]&Contrast$_0$&$r_1$ [mas]&PA$_1$ [deg.]&Contrast$_1$&$r_2$ [mas]&PA$_2$ [deg.]&Contrast$_2$}
  \startdata
   1PT &$76.0^{+0.6}_{-0.6}$&$340^{+1}_{-1}$& $248^{+8}_{-7}$& & &\\ 
   2PT &$74.3^{+1}_{-1}$&$324^{+1}_{-1}$&$288^{+14}_{-12}$ &$80.2^{+0.9}_{-0.8}$&$357^{+1}_{-2}$&$294^{+14}_{-13}$\\
   3PT &$74.0^{+1.2}_{-1.2}$&$315^{+3}_{-3}$&$418^{+42}_{-33}$ &$77.5^{+0.9}_{-0.9}$&$348^{+2}_{-2}$&$254^{+13}_{-10}$&$90.1^{+2.2}_{-2.5}$&$21^{+3}_{-2}$&$583^{+71}_{-58}$\\
   \hline
   2PG+IG+PT$^a$&$144^{+10}_{-14}$&$78^{+142}_{-2}$&$1292^{+330}_{-229}$& & &\\
   2PG+IG+PT$^b$&$143^{+7}_{-6}$&$76.3^{+1.4}_{-1.3}$&$1214^{+316}_{-215}$& $94^{+8}_{-5} /150^{+5}_{-6} $&$134^{+3}_{-3} / 135^{+1}_{-1} $&$1421^{+410}_{-295} / 1290^{+352}_{-212}$& $166^{+4}_{-4}$&$220^{+1}_{-1}$&$1323^{+447}_{-227}$ \\
   \enddata
\footnotesize{$^a$ Posterior is multi-modal, see Figure \ref{fig:plan_corn}.

$^b$ Extracted parameters from the most prominent modes found in the companion parameters of the 2PG+IG+PT fit, with $r_1$, PA$_1$ and Contrast$_1$ including the parameters of the two modes at PA $\sim$135 deg.}\\
\end{deluxetable*}  

\subsubsection{Rings and Planets}

To explore whether any point sources that were previously obscured by the bright disk emission are present when the disk emission is modelled, a point source component was added to the three component geometrical model (2PG+IG+PT), including inner and outer rings and a central disk. The best fit parameters were again extracted from the marginalized likelihood distributions obtained using dynamic nested sampling. To within the measured uncertainties, identical ring geometries are found as when the planetary point source is not included. Furthermore, the model including a planetary source has a somewhat higher evidence (Figure \ref{fig:all_like}).  

We note that this improvement in the fit is real, in that there is strong evidence for the planet model against the no planet model according to the computed natural logarithm of the Bayes factor of $\sim$16 $>$ 5 \citep{2008ConPh..49...71T}. Nevertheless, because our simple geometric models are unlikely to capture all of the expected real structure of the system, we can not rule out the possibility that such enhanced disk features are masquerading as planets in these fits. Thus, the (relatively) small increase in evidence should be treated with extreme caution. To further probe the residuals remaining in our data we also tested the effect of including alternative disk components, such as an additional uniform Gaussian ring, and observed similar small increases in the evidence. We note, however, that these various tests provided no consistency in the additional structure identified and thus we conclude that we have reached the limits of our geometrical model analysis.


\begin{figure*}
    \centering
    \includegraphics[width=1\textwidth]{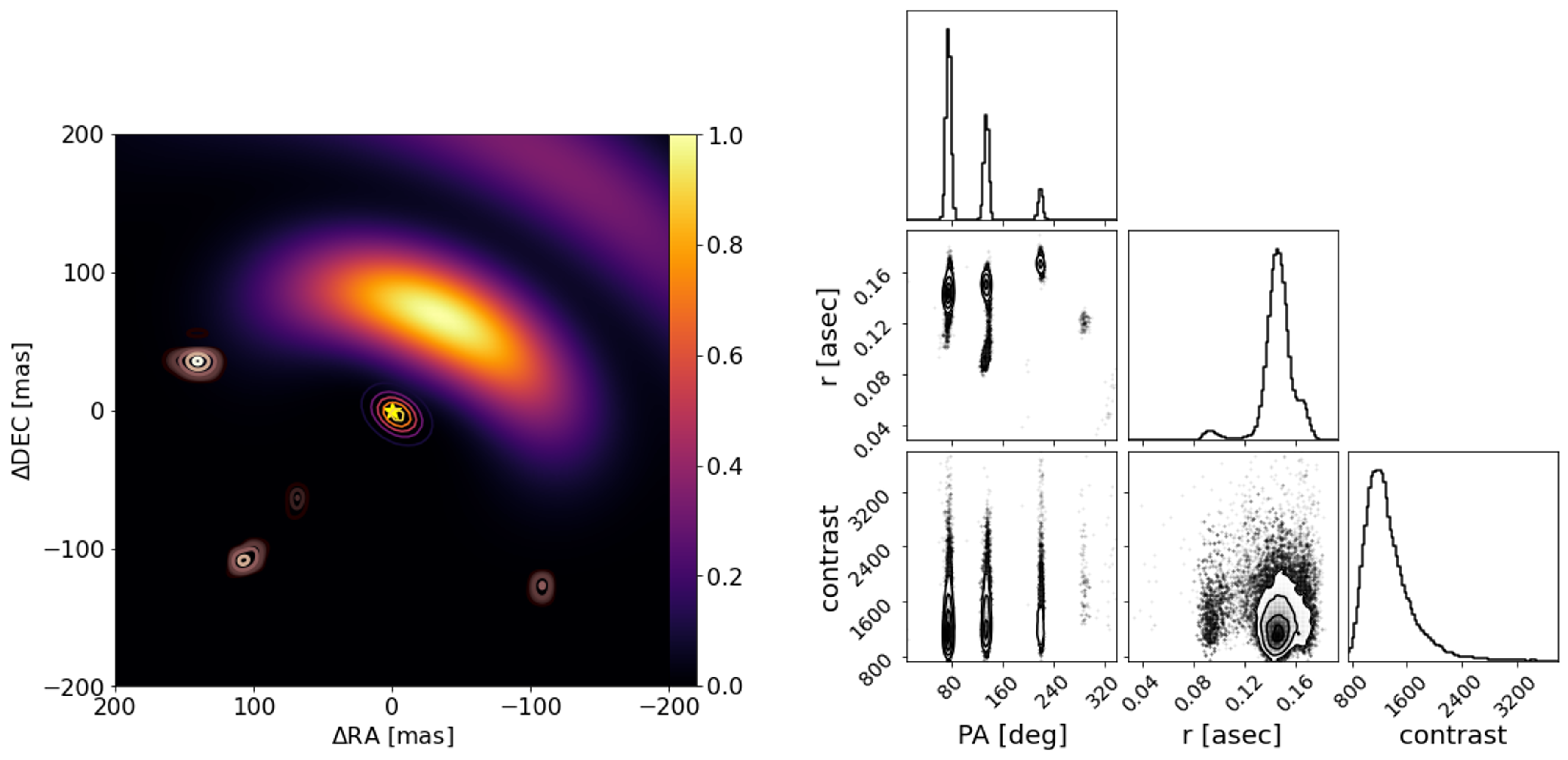}
    \caption{ 
    \textit{Left}: The joint posterior density (copper contours) of the RA and DEC position of a companion when it was fit for along with the two polar Gaussian ring plus inclined Gaussian model (background image + inclined Gaussian contours). Note that there are four potential locations for this planet found by the fitting. 
    \textit{Right}: Corner plot showing only the companion components from the model fit. The four most likely localized spots are clearly identifiable in the PA vs radius subplot.
    }
    \label{fig:plan_corn}
\end{figure*}

In Table \ref{comp} we show the extracted companion parameters from our best fit three component disk model plus planet (2PG+IG+PT). We note that the recovered companion brightness contrast is large, greater than 1000, and that the uncertainty in the contrast is $\sim$20\%. Furthermore, the marginalized distributions for the companion are multi-modal, so the quoted radius and position angle are ambiguous. 
{Thus, in the table we present also the parameters extracted from the most prominent modes. For illustrative purposes, in Figure \ref{fig:plan_corn} the best fit companion contours are plotted over our  2PG+IG model (left panel) along with a corner plot of the planet parameters (right panel).}

{From the modeling, it is clear that the rings and inner Gaussian account for the bulk of the asymmetric brightness recorded by the visibility measurements. This is in contrast to the models that only include point sources, where in order to fit the observations, the point source contrasts have to be much smaller, $\sim$250 for the brightest peak, and the uncertainties on the contrasts much lower (see Table \ref{comp}). The lack of a specific best-fit location for the candidate planet within a three component model (2PG+IG+PT), coupled with the large required brightness contrast and only marginal improvement in the evidence (Figure \ref{fig:all_like}), further emphasizes the need to treat the companion result with extreme caution. While the modeling does not rule out the presence of planets, it does limit their potential near infrared brightness. A broader, multi-band, and multi-epoch, study as well as further analysis of potential sources of disk asymmetry, beyond the polar Gaussian model, will be required to test against potential faint companion models and determine the robustness of these potential planet detections. 


\section{Conclusions}
\label{sec:conclusions}

We have analyzed VLT/SPHERE-IRDIS sparse aperture masking interferometry (SAM) data of the transition disk LkCa\,15 and fit geometrical models to our 2.1 and 2.3 $\mu$m data simultaneously, uncovering two previously identified arc components along with a marginally resolved, previously inferred, compact disk surrounding LkCa\,15. 
The robust 32 mas FWHM (5\,au) compact disk detection is an excellent demonstration of the power of SAM for imaging at and beyond the diffraction limit of a telescope. Furthermore, when using SAM, we demonstrate the importance of properly fitting extended structure before searching for candidate planets within transition disks. Our investigation has increased the lower limit contrast for candidate planets in the LkCa\,15 system by greater than a factor of three. 

We emphasize that for our model fitting approach, the advantage of geometrical models over radiative transfer models is that we can evaluate millions of models in minutes, whereas each radiative transfer model can take minutes to compute. Using radiative transfer informed models could be a next step to analyzing more realistic sources of asymmetry in transition disks, potentially using an emulator or accelerating current options using modern machine learning frameworks such as JAX \citep{jax2018github}}.

The main findings of our LkCa\,15 investigation are as follows:

\begin{itemize}
\item {Near infrared emission from the LkCa\,15 transition disk system is dominated by a series of arcs, spanning the spatial scales 5 - 50\,au recoverable via VLT SAM (Figure \ref{6_panel} and  Table \ref{2pg_par_bf}). These features, the larger of which were previously identified by near infrared direct imaging \citep{2014A&A...566A..51T,2015ApJ...808L..41T,2016ApJ...828L..17T,10.1093/pasj/psv133,Currie_2019}, lie inside a similar series of rings on scales 50 - 100\,au revealed in the mm using ALMA by \citet{2020A&A...639A.121F} (Figure \ref{fig:cont_over_curr});}

\item {A marginally resolved compact central Gaussian disk with FWHM $\sim$5\,au and slightly offset from LkCa\,15 is robustly detected by the VLT SAM observations. This component, which has previously been inferred \citep{Espaillat_2007,2010A&A...512A..11M,2010ApJ...717..441E,2018A&A...620A.195A} as belonging to LkCa\,15, demonstrates the power of SAM at small angular scales;}

\item {When companions are fit to the VLT SAM observations without first considering the extended disk structure, candidate planets are uncovered (Figure \ref{pt_sources}) with similar properties to those claimed previously \citep{Kraus_2011,Sallum_2015}. However, the companion fits are significantly poorer (lower evidence, see Figure \ref{fig:all_like}) than the extended disk structure results leading us to prefer the latter models;} 

\item {When companions are fit to the VLT SAM observations after including the extended disk structure, no clear evidence for candidate planets is recovered (Figure \ref{fig:plan_corn}). The near infrared contrast threshold for planets within 50\,au of LkCa\,15 is found to be at least $1000$ (Table \ref{comp}).}
\end{itemize}
\vfill

\section{acknowledgements}

\noindent We thank the anonymous referee for the insightful comments and feedback. We thank Thomas Vandal and Benjamin Pope for their helpful comments and suggestions. 
We acknowledge support from the Australian Research Council (DP 180103408) that funded part of this work (AMICAL). We also thank the ESO Paranal staff for support for conducting the observations reported in this paper. D.J.\ is supported by NRC Canada and by an NSERC Discovery Grant. J.S.B.\ acknowledges the support received from the UNAM PAPIIT project IA 101220 and from the CONACyT ``Ciencia de Frontera" project 263975.

\software{Tensorflow \citep{tensorflow2015-whitepaper}, dynesty \citep{Speagle_2020}, 
corner \citep{corner},   
Matplotlib \citep{Hunter:2007}, AMICAL \citep{2020SPIE11446E..11S}, Astropy \citep{2018AJ....156..123A}.
}




\bibliography{sample631}{}
\bibliographystyle{aasjournal}

\appendix

\section{Direct Image Reconstruction}
\label{app:dir}

\begin{figure*}[b]
\centering
    \includegraphics[width=0.95\textwidth]{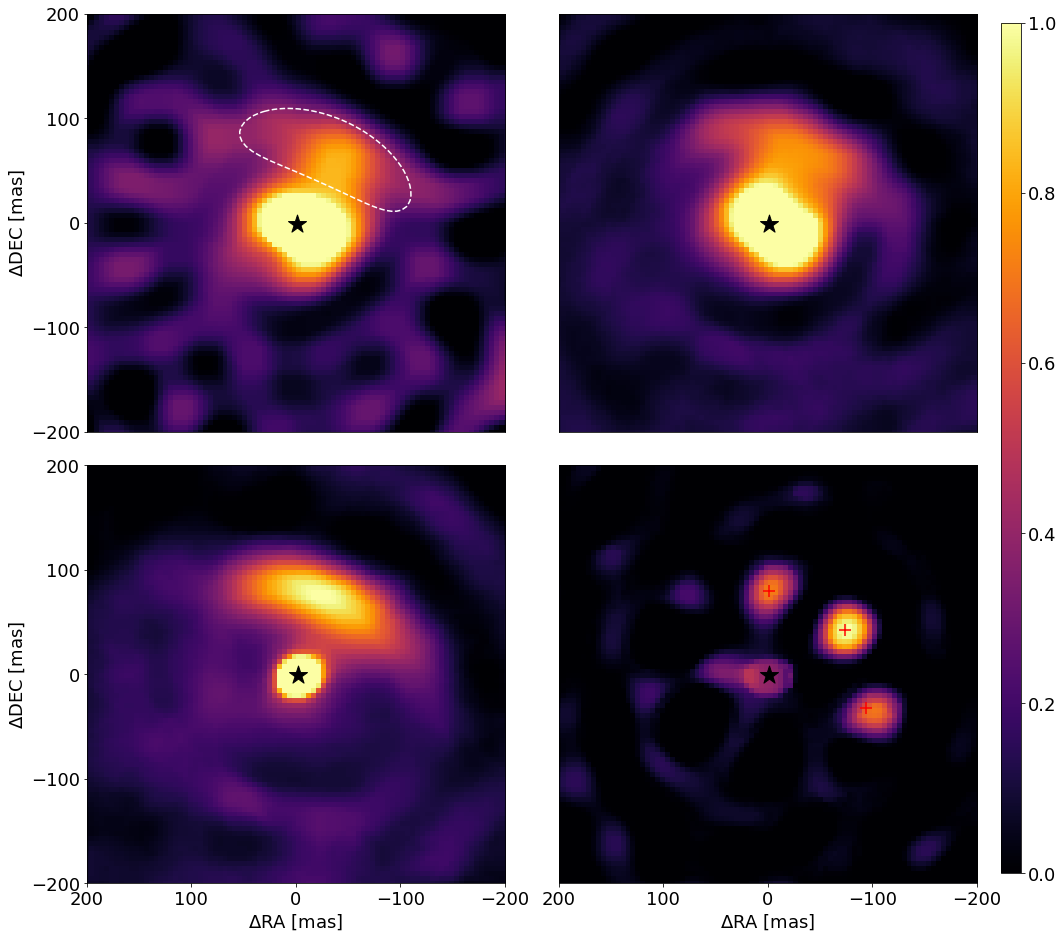}
    \caption{Median reconstructed images using our custom direct image reconstruction method (Section \ref{sec:imre}). \textit{Top Left}: Image reconstructed using the SAM  measurements of LkCa\,15, with the white contour denoting the half maximum of the inner arc component of the 2PG + IG geometric model (see also Figure \ref{6_panel}).
    \textit{Top Right}: Image reconstructed using simulated measurements from the best fit 2PG + IG geometric model. Note the similarity with the top left panel at the 
    arc location and in central asymmetry. 
    \textit{Bottom Left}: Image reconstructed using simulated measurements from the best fit 2PG + IG geometric model omitting the inner Gaussian component. Note the similarity in the 
    arc location with the top two panels and the dissimilarity in the central structure.  
    \textit{Bottom Right}: Image reconstructed using LkCa\,15 bcd point source model simulated measurements, applying contrasts from \citet{Sallum_2015} and predicted locations (denoted by red crosses) as shown in Figure \ref{pt_sources}. The  images in each panel are re-scaled such that the brightest pixel outside 45 mas from the central source is unity.}
    \label{direct}
\end{figure*}

The direct image reconstruction technique described in \S\,\ref{sec:imre}, along with a smooth starting image, was used to build an image of LkCa\,15 (Figure \ref{direct}) using the squared visibilities and closure phases. The use of this basic technique demonstrates that from an initial assumed brightness distribution consisting of a high contrast, central point source surrounded by an extended symmetric Gaussian brightness distribution, the most likely reconstruction includes an extended arc to the northwest of the origin along with marginally resolved {asymmetric} emission surrounding the central point source. Convergence to the arc solution required 250 iterations.  The contrasts used for the initial images were randomly generated with a mean star to total Gaussian brightness contrast of roughly 2 and with the spatial extent of the Gaussian being randomly selected from a distribution with mean FWHM $\sim$850 mas. This process was repeated for 10 randomly generated initial contrasts and Gaussian sizes for the input initial maps. The final reconstructed image shown in the top left panel of Figure \ref{direct} is the median of the results of 10 individual runs of 250 steps each. We note that in order to converge, it is necessary to begin with a high contrast between the central point source and the extended Gaussian. 

{The same direct reconstruction procedure was repeated to build images using simulated data from our best fit 2PG + IG model (top right panel in Figure \ref{direct}), our best fit 2PG + IG model omitting the inner compact Gaussian component (bottom left panel in Figure \ref{direct}), as well as for a LkCa\,15 bcd point source model (bottom right panel in Figure \ref{direct}). These tests allow us to probe the significance of any particular recovered structure as a function of the input data set. In each case, randomly generated Gaussian noise on the level expected to be present in our data was added to the simulated data before each of the 10 resets. The similarity in arc and central asymmetric structure observed for both the reconstructed image from the SAM measurements and from the simulated data measurements of the best fit 2PG + IG model, as well as both of their obvious dissimilarity with the reconstructed image of the point source simulated data measurements supports the validity of our geometrical modelling. Additionally, the obvious dissimilarity between the asymmetric, elliptical central emission seen in the reconstructed images for the SAM data and the 2PG + IG model compared to the symmetric central emission seen in the reconstruction for the 2PG + IG model omitting the inner compact Gaussian component further validates the inclusion of the inner Gaussian component used in our geometrical modelling.} 

We note that the outer ring is not at all constrained during these direct image reconstructions. Image reconstruction is an ill-posed inverse problem, with in our case 65536 free parameters and only 1736 data measurements. We thus only expect to be able to recover the most prominent features in our data, including the inner arc and, potentially, the central, Gaussian disk. The outer ring is near the outer edge of the spatial scales to which our data are sensitive and appears fainter than the two inner components. We therefore expect that for it to be recovered using direct image reconstruction strong priors would have to be imposed.

\end{document}